\documentclass[runningheads]{svproc}

\usepackage{amsmath}
\usepackage{graphicx}
\usepackage{easyReview}

\usepackage{booktabs}
\newcommand{\mr}[1]{\mathrm{#1}}

\begin{document}

\title{Driveability Constrained Models for Optimal Control of Hybrid Electric Vehicles}
\titlerunning{\emph{Preprint submitted to the 2nd IFToMM for SDG Workshop}}

\author{Federico Miretti \and
	Daniela Misul}

\authorrunning{F. Miretti and D. Misul}

\institute{CARS@Polito - Center for Automotive Research and Sustainable Mobility, Politecnico di Torino, C.so Ferrucci 112, Torino, TO 10138, Italy\\
	\email{federico.miretti@polito.it}\\
	\url{www.cars.polito.it}}

\maketitle
	
\begin{abstract}
	This work investigates the effect of three different driveability constraints on the optimal energy management strategy for a p2 parallel hybrid. Two of these constraints are used to prevent frequent gear shifting and engine start/stops, while the third is used to increase the sportiness of the vehicle by maximizing the available torque reserve at	all times.
	The constraints are imposed by reformulating them as penalty terms to be added to the base running cost of the control strategy, which is fuel consumption. Dynamic programming, a popular optimal control technique, is then used to design the energy management strategy that minimizes the total cost. A case study is developed for a p2 parallel hybrid and simulated on a combination of the Artemis driving cycles.
	The impact of each driveability constraint is analyzed with respect to a set of relevant features of the control strategy, such as the choice of engine	operating points and the gear shift pattern. The resulting discussion provides some useful insight for the design of real-time, rule-based control strategies.
\end{abstract}

\section{Introduction}
Hybrid electric vehicles (HEVs) enable fuel economy improvement by exploiting the additional degree of freedom granted by the presence of an electrical power source to operate the thermal engine at higher efficiencies. A dedicated controller, often called the energy management system (EMS), is needed to define how these two power sources are used to meet the driver's power demand. Additionally, since HEVs generally employ automated transmissions, its control is also included in the EMS~\cite{Guzzella2007,Onori2015}.
The EMS has a strong influence on the powertrain's performance in terms of fuel economy, emissions and driveability. Hence, a great deal of attention has been devoted to EMS design both within the industry and in academic research, and a wide range of techniques has been proposed.

Among these, dynamic programming is one of the most popular. Because of its flexibility and guaranteed optimality, it can be easily and effectively used to analyze optimal control strategies in the design phase, in an off-line simulation environment. Its most notable drawbacks are a general inability to deal with complex simulation models because of its computational burden and the need for advance knowledge of the vehicle's speed profile in time. The latter in particular makes the technique unsuitable for real-time control.

Still, dynamic programming can be highly effective in supporting the design of real-time control strategies. For example, rule-based controllers can be designed by applying some rule extraction procedure to the control trajectories generated by dynamic programming~\cite{Lin2001,Peng2017,Mansour2016}, or the same results can be used to calibrate some other optimization\-based EMS~\cite{Anselma2022b}.
Unfortunately, when used to derive fuel-optimal control strategies,  dynamic programming typically induce a number of undesirable driveability issues such as frequent engine start/stops, erratic gear shifting, and a general lack of sportiness. This is also typical of other common optimal control approaches such as equivalent consumption minimization strategies (ECMS) and Pontryagin's minimum principle (PMP).

Many works can be found in the literature which address some or all of these problems, which are commonly referred to as driveability issues.
Possibly the earliest of such applications can be found in~\cite{Lin2003}, where a penalty term associated with gear shifting was included in a dynamic programming algorithm whose results were then used to develop a rule-based control strategy.
Several authors used stochastic dynamic programming to reduce the frequency of gear shifts~\cite{Li2016}, engine starts~\cite{Leroy2012}, or both~\cite{Opila2012}, by adding corresponding penalty terms to the running cost in the cost functional.
Similarly, the authors in~\cite{Anselma2020,Anselma2022} embedded penalty terms for gear shifts and engine starts in an heuristic framework named SERCA and in a dynamic programming application to act as a benchmark.

Torque (or power) reserve appears to be the least considered among driveability aspects. To the best of our knowledge, only two works have tackled this issue, both of which employed some variant of the ECMS.
One approach for a multi-mode PHEV with heuristic penalty factors for each mode transition was developed in~\cite{MiroPadovani2016}, while also including a hard constraint for the available torque reserve at the wheels.
In contrast,~\cite{VidalNaquet2012} dealt with torque reserve by adding a penalty term to the equivalent consumption, in addition to a gear shift penalty.

In this work, we developed a four-term cost functional to be used in a dynamic programming framework, which includes fuel consumption, penalties for gear shifts and engine starts and a penalty term for the available torque reserve. We then developed a case study with a p2 parallel hybrid, whose main parameters can be found in Table~\ref{tab:vehicle_parameters}, and assessed the effect of each penalty term on the obtained control strategies. Finally, we discuss the implications of our results on the development of real-time heuristic control strategies. 

\begin{table}[htbp]
	\caption{Main vehicle data.}
	\label{tab:vehicle_parameters}
	\centering
	\begin{tabular}{@{}lll@{}} \toprule
		Component & Parameter & Value \\ \midrule
		Vehicle & Mass & 1300 kg \\
		& First coast-down coefficient & 150 N \\
		& Second coast-down coefficient & 2.24 N/(m\,\textperiodcentered\,s) \\
		& Third coast-down coefficient & 0.44 N/(m\,\textperiodcentered\,s)\textsuperscript{2} \\
		& Tyre radius & 0.327 m \\
		Transmission & Gear ratios & [3.46, 1.844, 1.258, 1.027, 0.85] \\
		& Efficiency & [0.93, 0.94, 0.947, 0.948, 0.946] \\
		Engine & Displacement & 0.9 l \\	
		& Rated power & 52 kW \\
		& Maximum torque & 85 Nm \\
		E-machine & Rated power & 30 kW \\
		& Maximum torque & 200 Nm \\
		Battery & Type & Li-ion \\
		& Nominal capacity & 5.3 Ah	\\
		& Nominal voltage & 295 V	\\
		\bottomrule
	\end{tabular}
\end{table}

\section{Simulation model}
The simulation model was developed using a backward-facing approach~\cite{Guzzella2007,Onori2015}, as is typical for control-oriented models in EMS design.
The tractive effort $F_\mathrm{veh}$ was evaluated with a simple longitudinal model considering the resistant forces $F_\mathrm{res}$ (using road load coefficients $k_0$, $k_1$ and $k_2$) and the vehicle's inertia
\begin{equation}
	F_\mathrm{veh} = F_\mathrm{res} + m_\mr{veh} a_\mathrm{veh} = k_0 + k_1 v_\mathrm{veh} + k_2 v_\mathrm{veh}^2 + m_\mr{veh} a_\mathrm{veh}.
\end{equation}

A quasi-static powertrain model was then used to propagate this tractive effort through the wheels, final drive and gearbox to obtain a torque demand $T_\mathrm{d}$, which for the p2 hybrid considered in this work refers to the gearbox input.
\begin{equation}
	\label{eq:trqDemand}
	T_\mathrm{d} = \frac{F_\mathrm{veh} r_\mathrm{wh}}{\tau_\mathrm{fd} \tau_\mathrm{gb}(\gamma)},
\end{equation}
Here, $r_\mathrm{wh}$ is the wheel radius, $\tau_\mathrm{fd}$ and $\tau_\mathrm{gb}$ are the final drive and gearbox speed ratios, and $\gamma$ represents the gear number.
This torque demand was then split between the engine and the e-machine based the torque-split factor $\alpha_\mr{eng}$:
\begin{equation}
	\alpha_\mr{eng} = \frac{T_\mr{eng}}{T_\mr{d}}.
\end{equation}

The engine and e-machine were characterized by a steady-state fuel flow rate map $\dot{m}_\mathrm{f}(\omega_\mr{eng}, T_\mr{eng})$ and an efficiency map $\eta_\mathrm{em}(\omega_\mr{em}, T_\mr{em})$, as well as torque limit curves and speed constraints.
The e-machine efficiency was used to evaluate the battery electrical power $P_\mathrm{b}$. The battery current $i_b$ was evaluated as a function of the battery power $P_\mathrm{b}$ with an equivalent circuit model:
\begin{equation}
	\label{eq:battModel}
	P_\mathrm{b} = v_b i_b = \left( v_\mathrm{oc}(\sigma) + R_0(\sigma) i_\mathrm{b} \right) \, i_b.
\end{equation}
where $\sigma$ is the battery state of charge and $v_\mathrm{oc}(\sigma)$ and $R_0(\sigma)$ are the open-circuit voltage and internal resistance characteristics. 

\section{EMS design with dynamic programming}
Dynamic programming is an optimal control technique to control the evolution of a dynamical system in time while minimizing some additive cost $J$.
In the context of dynamic programming, the simulation is discretized in $N$ time steps. The model is characterized by a set of control variables $u$ which influence the system's state evolution as defined by the state dynamics $x_{k+1} = f(x_k, u_k, w_k)$ while incurring in a total cost $J(x_0) = \sum_k^{N-1} L(x_k, u_k, w_k)$, where $L$ is the stage cost. The exogenous input $w_k$ is used to characterize the set of variables which affect the simulation without being influenced by the controls; in powertrain simulation models, they are generally identified with the speed and acceleration profiles of the prescribed driving mission.

The model that was developed for this work uses three state variables to characterize the battery's state of charge $\sigma$, the gear number for the previous time step $\gamma_\mr{p}$ and the engine state for the previous time step $\epsilon_\mr{p}$, i.e.
\begin{equation}
	x = \begin{pmatrix} \sigma \\ \gamma_\mr{p} \\ \epsilon_\mr{p} \end{pmatrix},
\end{equation}
and two control variables to set the engine torque-split factor $\alpha_\mr{eng}$ and the gear number for the current time step $\gamma$, i.e.
\begin{equation}
	u = \begin{pmatrix} \alpha_\mr{eng} \\ \gamma \end{pmatrix}.
\end{equation}
The engine torque-split ratio was selected to characterize the powerflow over other common choices as it highly interpretable, i.e. there is a direct correspondence between the value of $\alpha_\mr{eng}$ and the operating mode~\cite{Miretti2022}.

The running cost was set to a trade-off of four different terms:
\begin{equation}
	\label{eq:costfun}
	L = \dot{m}_\mr{fuel} \, \Delta t + L_\gamma + L_\epsilon + L_{T_\mr{res}}.
\end{equation}
The first term is the fuel consumption over a time step $ \Delta t$ ($\dot{m}_\mr{fuel}$ being the fuel flow rate), so that the fuel consumption over the whole mission will be minimized. The remaining three terms $L_\gamma$,  $L_\epsilon$, $L_{T_\mr{res}}$ are penalty terms that penalize gear shifting, engine starts and low torque reserve availability respectively.

The gear shift penalty was defined by a factor $\phi_{\gamma}$ which is applied each time a gear shift occurs:
\begin{equation}
	L_{\gamma} =
	\begin{cases}
		\phi_{\gamma} & \text{if } \gamma \neq \gamma_\mr{p},\\
		0 & \text{otherwise.}
	\end{cases}
\end{equation}
Similarly, the engine start penalty was defined by a factor $\phi_{\epsilon}$ which is applied each time the engine is turned on. An engine start occurs when a non-zero torque is set and the engine was off at the previous time step:
\begin{equation}
	L_{\gamma} =
	\begin{cases}
		\phi_{\epsilon} & \text{if } \alpha_\mr{eng} > 0 \wedge \epsilon_\mr{p} = 0,\\
		0 & \text{otherwise}.
	\end{cases}
\end{equation}

The torque reserve penalty was defined as the ratio between the used powertrain torque $T_\mr{pwt}$ and the available powertrain torque $T_\mr{pwt,max}$, multiplied by a tunable factor $\phi_{T_\mr{res}}$. The penalty is only applied if the vehicle is neither braking nor at standstill:
\begin{equation}
	L_{T_\mr{res}} =
	\begin{cases}
		\phi_{T_\mr{res}} \cdot \frac{T_\mr{pwt}}{T_\mr{pwt,max}} & \text{if } T_\mr{req} > 0 \wedge v_\mr{veh} > 0,\\
		0 & \text{otherwise}.
	\end{cases}
\end{equation}
More specifically, the powertrain torque is defined as the sum of the engine and e-machine torque at the gearbox input:
\begin{equation}
	T_\mr{pwt} = T_\mr{eng} + \max{(T_\mr{em} \tau_\mr{tc}, 0)}.
\end{equation}
Note that the e-machine torque was subject to lower saturation at zero in order to prevent its torque in generator mode being counted while using the powertrain in battery charging mode.
Finally, available powertrain torque $T_\mr{pwt,max}$ was simply defined as
\begin{equation}
	T_\mr{pwt,max} = T_\mr{eng,max} + T_\mr{em,max} \tau_\mr{tc}.
\end{equation}
Since both the engine and e-machine maximum torque are dependent on their speed, they are influenced by the gear engaged in the gearbox. Hence, the torque reserve penalty can be affected by the EMS by changing the gear number.

\section{Case study}
\begin{figure}
	\includegraphics{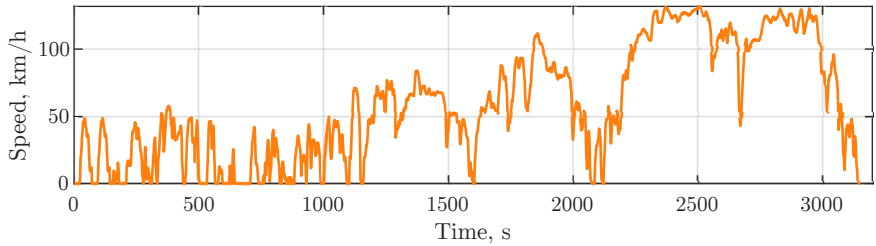}
	\caption{The simulated driving cycle.}
	\label{fig:artemis}
\end{figure}

In order to assess the effect of driveability constraints on the fuel-optimal control strategy, we implemented the simulation model described in the previous section in MATLAB and we used a dedicated dynamic programming solver called DynaProg~\cite{Miretti2021} to obtain optimal control strategies with the cost functional formulated in Eq.~\ref{eq:costfun}. For the driving cycle, a combination of the Artemis Urban, Artemis Rural Road and Artemis Motorway 130 cycles was used as shown in Fig.~\ref{fig:artemis}, with a total length of 51 kilometers and duration of 52 minutes.

With this framework, we developed four different cases by tuning the cost functional. In the first case, we set all driveability penalties to zero, considering fuel economy only as our objective. In the remaining three cases, we considered fuel economy and one driveability penalty at a time, disregarding the other two.
In the remainder of this section, these strategies will be referred to as:
\begin{enumerate}
	\item[a)] fuel-optimal: no penalty terms for driveability are considered.
	\item[b)] gear shift-penalty: fuel-optimal with a penalty term for gear shifting.
	\item[c)] engine start penalty: fuel-optimal with a penalty term for engine starts.
	\item[d)] torque reserve penalty: fuel-optimal with a penalty term for torque reserve.
\end{enumerate}
The fuel-optimal strategy produced a fuel economy of 4.58 l/100km, an average of 18 gear shifts per minute and 3.1 engine starts per minute, with an average torque reserve of 58.3 \%.
The penalty factors for the other strategies were tuned with three separate parameter sweeps to obtain a sensible trade-off between fuel economy and each driveability objective. In particular, we aimed at less than one gear shift per minute for strategy b), less than 0.67 engine starts for strategy c) and an average torque reserve of at least 65 \%
for strategy d). The corresponding fuel consumption increase for each strategy is reported in Table~\ref{tab:mainres}.

\begin{table}[htbp]
	\caption{Performance of the four strategies.}
	\label{tab:mainres}
	\centering
	\begin{tabular}{@{}lllll@{}} \toprule
		& Fuel economy & Gear shifts & Engine starts & Torque reserve \\ 
		&  & \#/min & \#/min & \% \\
		\midrule
		a) fuel-optimal & 4.58 l/100km & 18 & 3.1 & 58.3 \% \\
		b) gear shift-penalty & +1.6 \%  & 0.93 & 2.7 & 60.7 \% \\
		c) engine start penalty & +3.2 \% & 14  & 0.67  & 55.6 \% \\
		d) torque reserve penalty & +2.5 \% & 16.4  & 3.3  & 65.8 \% \\
		\bottomrule
	\end{tabular}
\end{table}

\begin{figure}[p]
	\centering
	\includegraphics{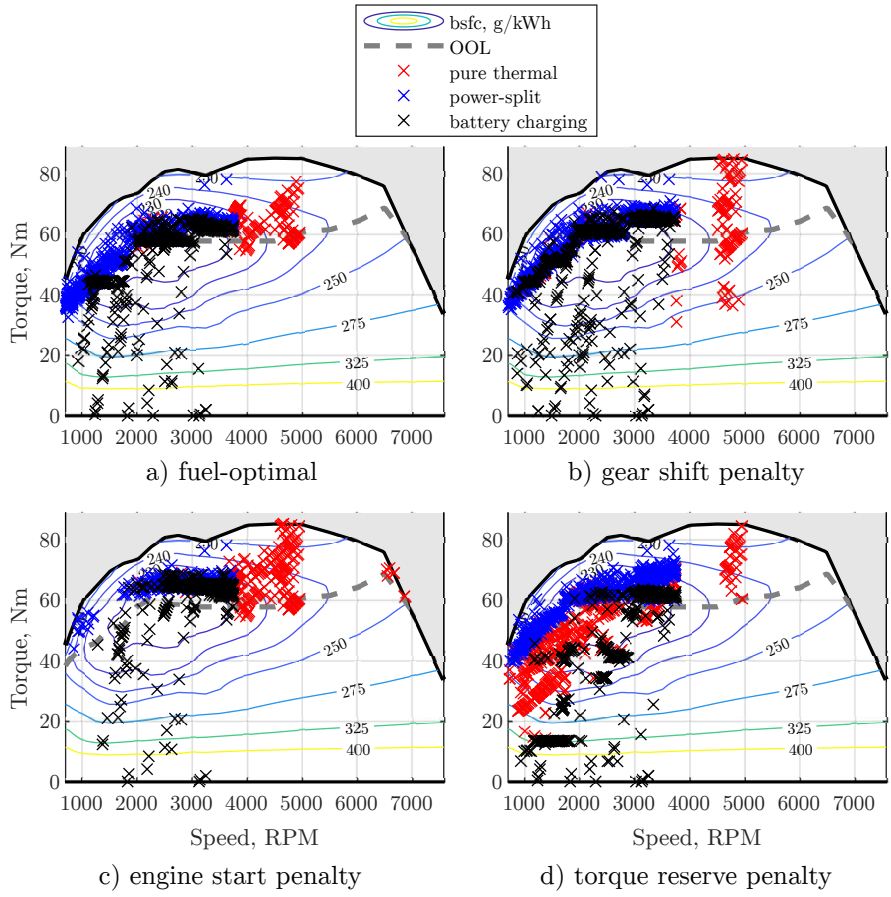}
	\caption{Comparison of engine operating maps with four different cost functions.}
	\label{fig:engMapComparison}
\end{figure}

Fig.~\ref{fig:engMapComparison} shows the engine operating points throughout the mission for the four different strategies, color-coded based on the adopted operating mode.
As expected, the engine tends to work near the optimal operating line (OOL) for the fuel-optimal control strategy. Introducing the gear shift penalty in b), the most notable difference is that the pure thermal operating points are now concentrated into two distinct and narrower speed ranges. These points are operated with the third and fourth gear engaged; clearly, the unconstrained strategy in a) uses frequent shifting between this two to move more points closer to the OOL.

\begin{figure}[p]
	\centering
	\includegraphics{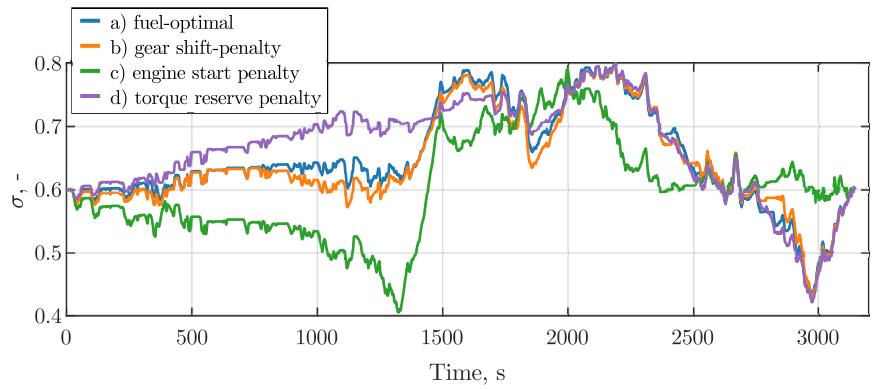}
	\caption{Comparison of the battery state of charge profile with the four strategies.}
	\label{fig:soc}
\end{figure}

Considering the engine start penalty in c), we can note an increased usage of the pure thermal mode and a decrease in the usage of power-split mode, which is also evident from Table~\ref{tab:timeshares}.
In particular, this strategy makes a wider use of pure electric mode during the Urban phase of the driving cycle, discharging the battery, and uses the Rural Road phase to charge the battery back up; this is clearly visible from the state of charge profiles in Fig.~\ref{fig:soc}.
Still, the areas where the engine operating points concentrate remain similar.

\begin{table}[htbp]
	\caption{Time shares spent in each operating mode with the four strategies.}
	\label{tab:timeshares}
	\centering
	\begin{tabular}{@{}lllll@{}} \toprule
		& Pure electric & Pure thermal & Power-split & Battery charging \\ \midrule
		a) fuel-optimal & 42.8 \% & 7.91 \% & 26.2 \% & 23.1 \% \\
		b) gear shift-penalty & 43.3 \% & 6.64 \% & 25.9 \% & 24.1 \% \\
		c) engine start penalty & 61.8 \% & 11.4 \% & 8.42 \% & 18.4 \% \\
		d) torque reserve penalty & 40.4 \% & 17.6 \% & 24.5 \% & 17.5 \% \\
		\bottomrule
	\end{tabular}
\end{table}

Finally, the effect of the torque reserve penalty in d) generates a large number of pure thermal points in the low-speed region of the map. These points are all points that provide a good trade-off between fuel economy and sportiness, because they are concentrated along the OOL and at the same time they leave the full torque of the e-machine, in its constant torque region, available.

\begin{figure}[h]
	\centering
	\includegraphics{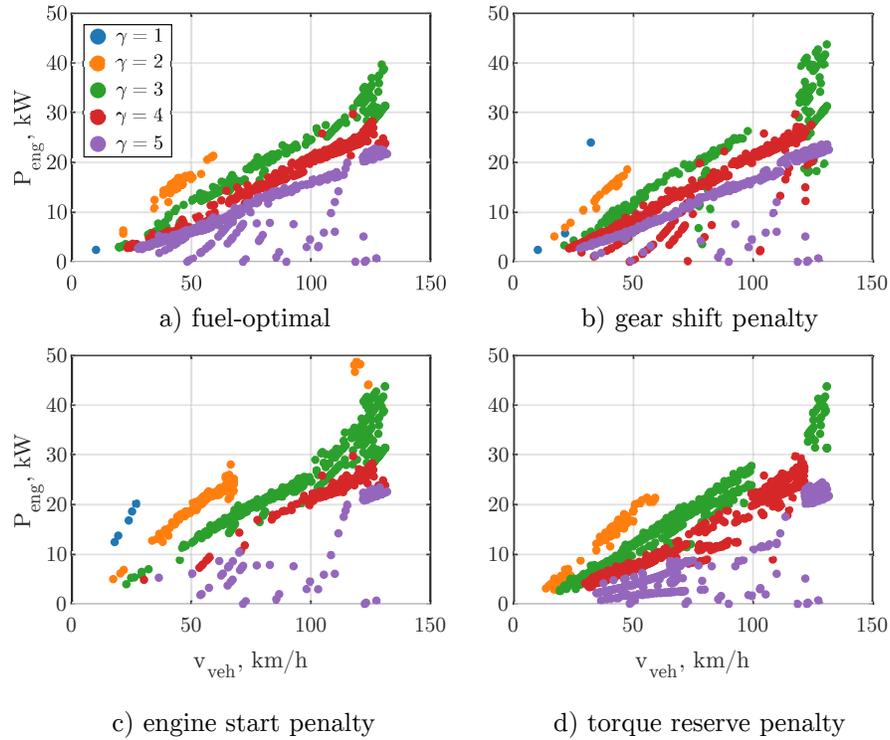}
	\caption{Comparison of gear shifting patterns with the four strategies. Only hybrid operating modes are represented (pure electric points are not shown).}
	\label{fig:gearsComparison}
\end{figure}

We now turn our attention to gear shift behavior in Fig.~\ref{fig:gearsComparison}, which shows how the engaged gears relate to the vehicle speed and engine power; this is a typical analysis tool when designing gear shift schedules for automated transmissions. Note that only hybrid modes are represented, i.e. pure electric points are not depicted. 

Considering the fuel-optimal strategy in a), we observe that a clear shifting pattern emerges as the operating points are neatly separated based on the engaged gear. We also note that the first gear is almost never engaged, as low speed operation is driven almost exclusively in pure electric. 

Introducing a gear shift penalty in b), however, complicates the shifting behavior. Although it is still possible to identify preferred areas for each gear, there are significant overlays such as the third and fourth gear being engaged in the area previously reserved to the fifth gear at several speeds. This is likely a consequence of the strategy having to sometimes operate in a non-efficient way in order to limit the number of gear shifts. 

The strategy with a penalty for engine starts in c) instead shows a more regular shifting pattern; the most notable difference with respect to the fuel-optimal strategy is a reduced usage of the fifth gear, which is mostly engaged at high power; further inspection revealed that these points were engaged in battery charging mode. Also noticeable is an increased usage of the third gear at higher power; these correspond to the additional pure thermal operating points.

Finally, introducing the torque reserve penalty in d) generated a larger concentration of operating points at high power and high speed for the fourth and fifth gear, which correspond to the additional pure thermal and battery charging points that we previously observed in Fig.~\ref{fig:engMapComparison}.

\section{Conclusions}
In this work, we implemented dynamic programming to investigate the effect of three different driveability constraints on the optimal energy management strategy for a p2 parallel hybrid. The constraints were implemented by adding three different penalty terms to the base cost of the optimal control problem, which is fuel consumption.

By testing each penalty term individually, we were able to assess the impact of each corresponding driveability aspect on a set of relevant features of the control strategy, such as the choice of engine operating points and the gear shift pattern. These considerations provide useful insight for the development of real-time, rule-based control strategy that minimize fuel consumption while preventing unrealistic and potentially damaging gear shifting and engine start/stop behavior, as well as targeting varying levels of sportiness.

\bibliographystyle{splncs04}
\bibliography{references}

\begin{thebibliography}{10}
\providecommand{\url}[1]{\texttt{#1}}
\providecommand{\urlprefix}{URL }
\providecommand{\doi}[1]{https://doi.org/#1}

\bibitem{Anselma2022}
Anselma, P.G.: Computationally efficient evaluation of fuel and electrical
  energy economy of plug-in hybrid electric vehicles with smooth driving
  constraints. Applied Energy  \textbf{307} (Feb 2022).
  \doi{10.1016/j.apenergy.2021.118247}

\bibitem{Anselma2020}
Anselma, P.G., Biswas, A., Belingardi, G., Emadi, A.: Rapid assessment of the
  fuel economy capability of parallel and series-parallel hybrid electric
  vehicles. Applied Energy  \textbf{275} (Oct 2020).
  \doi{10.1016/j.apenergy.2020.115319}

\bibitem{Anselma2022b}
Anselma, P.G., Spano, M., Capello, M., Misul, D., Belingardi, G.: Calibrating a
  real-time energy management for a heavy-duty fuel cell electrified truck
  towards improved hydrogen economy. In: {SAE} Technical Paper Series. {SAE}
  International (jun 2022). \doi{10.4271/2022-37-0014}

\bibitem{Guzzella2007}
Guzzella, L., Sciarretta, A.: Vehicle Propulsion Systems Introduction to
  Modeling and Optimization. Springer London, Limited (2007)

\bibitem{Leroy2012}
Leroy, T., Malaizé, J., Corde, G.: Towards real-time optimal energy management
  of {HEV} powertrains using stochastic dynamic programming. In: 2012 {IEEE}
  {Vehicle} {Power} and {Propulsion} {Conference}. pp. 383--388 (Oct 2012).
  \doi{10.1109/VPPC.2012.6422661}, iSSN: 1938-8756

\bibitem{Li2016}
Li, L., Yan, B., Yang, C., Zhang, Y., Chen, Z., Jiang, G.:
  Application-{Oriented} {Stochastic} {Energy} {Management} for {Plug}-in
  {Hybrid} {Electric} {Bus} {With} {AMT}. IEEE Transactions on Vehicular
  Technology  \textbf{65}(6),  4459--4470 (Jun 2016).
  \doi{10.1109/TVT.2015.2496975}

\bibitem{Lin2001}
Lin, C.C., Kang, J.M., Grizzle, J., Peng, H.: Energy management strategy for a
  parallel hybrid electric truck. In: Proceedings of the 2001 {American}
  {Control} {Conference}. ({Cat}. {No}.{01CH37148}). vol.~4, pp. 2878--2883
  (Jun 2001). \doi{10.1109/ACC.2001.946337}

\bibitem{Lin2003}
Lin, C.C., Peng, H., Grizzle, J., Kang, J.M.: Power management strategy for a
  parallel hybrid electric truck. IEEE Transactions on Control Systems
  Technology  \textbf{11}(6),  839--849 (Nov 2003).
  \doi{10.1109/TCST.2003.815606}

\bibitem{Mansour2016}
Mansour, C.J.: Trip-based optimization methodology for a rule-based energy
  management strategy using a global optimization routine: the case of the
  prius plug-in hybrid electric vehicle. Proceedings of the Institution of
  Mechanical Engineers, Part D: Journal of Automobile Engineering
  \textbf{230}(11),  1529--1545 (aug 2016). \doi{10.1177/0954407015616272}

\bibitem{Miretti2022}
Miretti, F., Misul, D.: Robust {Modeling} for {Optimal} {Control} of {Parallel}
  {Hybrids} {With} {Dynamic} {Programming}. In: 2022 {IEEE} {Transportation}
  {Electrification} {Conference} \& {Expo} ({ITEC}). pp. 1015--1020 (Jun 2022).
  \doi{10.1109/ITEC53557.2022.9813982}

\bibitem{Miretti2021}
Miretti, F., Misul, D., Spessa, E.: {DynaProg}: {Deterministic} {Dynamic}
  {Programming} solver for finite horizon multi-stage decision problems.
  SoftwareX  \textbf{14},  100690 (Jun 2021). \doi{10.1016/j.softx.2021.100690}

\bibitem{MiroPadovani2016}
Miro-Padovani, T., Colin, G., Ketfi-Chérif, A., Chamaillard, Y.:
  Implementation of an {Energy} {Management} {Strategy} for {Hybrid} {Electric}
  {Vehicles} {Including} {Drivability} {Constraints}. IEEE Transactions on
  Vehicular Technology  \textbf{65}(8),  5918--5929 (Aug 2016).
  \doi{10.1109/TVT.2015.2476820}

\bibitem{Onori2015}
Onori, S., Serrao, L., Rizzoni, G.: Hybrid Electric Vehicles Energy Management
  Strategies. Springer London, Limited (2015)

\bibitem{Opila2012}
Opila, D.F., Wang, X., McGee, R., Gillespie, R.B., Cook, J.A., Grizzle, J.W.:
  An {Energy} {Management} {Controller} to {Optimally} {Trade} {Off} {Fuel}
  {Economy} and {Drivability} for {Hybrid} {Vehicles}. IEEE Transactions on
  Control Systems Technology  \textbf{20}(6),  1490--1505 (Nov 2012).
  \doi{10.1109/TCST.2011.2168820}

\bibitem{Peng2017}
Peng, J., He, H., Xiong, R.: Rule based energy management strategy for a
  series–parallel plug-in hybrid electric bus optimized by dynamic
  programming. Applied Energy  \textbf{185},  1633--1643 (Jan 2017).
  \doi{10.1016/j.apenergy.2015.12.031}

\bibitem{VidalNaquet2012}
Vidal-Naquet, F., Zito, G.: Adapted optimal energy management strategy for
  drivability. In: 2012 {IEEE} {Vehicle} {Power} and {Propulsion} {Conference}.
  pp. 358--363 (Oct 2012). \doi{10.1109/VPPC.2012.6422678}

\end{thebibliography}

\end{document}